\documentclass[aps,prd,preprint,showpacs,superscriptaddress]{revtex4}
\usepackage{graphicx}
\newcommand{\U}{{\cal U}}
\begin{document}
\draft
\title{Unparticle physics in top pair signals at the LHC and ILC}
\author{A.T. Alan}
\email[ alan\_a@ibu.edu.tr]{}
 \affiliation{Department of
Physics, Abant Izzet Baysal University, 14280 Bolu, Turkey}
\author{N.K. Pak}
\email[pak@metu.edu.tr]{}
 \affiliation{Department of Physics,
Middle East Technical University, 06531 Ankara, Turkey}
\begin{abstract}
We study the effects of unparticle physics in the pair productions
of top quarks at the LHC and ILC. By considering vector, tensor and
scalar unparticle operators, as appropriate, we compute the total
cross sections for pair production processes depending on scale
dimension $d_{\U}$. We find that the existence of unparticles would
lead to measurable enhancements on the SM predictions at the LHC. In
the case of ILC this may become two orders of magnitude larger than
that of SM, for smaller values of $d_\U$, a very striking signal for
unparticles.
\end{abstract}
\maketitle
\section{introduction}
Scale invariance is broken in the quantum field theory by the masses
of particles. Recently, Georgi \cite{Georgi:2007ek,Georgi:2007si}
has introduced a new scheme based on the existence of a non-trivial
scale invariant sector at a very large scale $M_\U$. The fields of
this sector and Standard Model (SM) fields can interact via the
exchange of a connector stuff. Below this mass scale,
non-renormalizable operators are suppressed by power of $1/M_{\U}$.
Renormalization effects of the scale invariant Banks and Zaks
($\mathcal{BZ}$) sector \cite{Banks:1981nn} induces dimensional
transmutation at the scale $\Lambda_{\U}$. Below the scale
$\Lambda_{\U}$, $\mathcal{BZ}$ operators transform to so-called
unparticle operators with scale dimension $d_{\U}$. So far there
have been many studies exploring various aspects of unparticles,
including also supersymmetric extensions and colored versions
\cite{Cheung:2007zz}.

Top quark physics is a very interesting and active field of research
\cite{Beneke:2000hk,Wagner:2005jh}. One of the most distinctive
features of top quark which makes it so interesting is its large
mass. Due to this large mass the top quark is considered as an ideal
tool for probing new physics beyond SM. The CERN Large Hadron
Collider (LHC) will be, in a sense, a top quark factory with about
$10^7$ top pair signals per year. This large statistic will also
enable us to determine the top quark properties very accurately, at
the LHC. On the other hand, although the cross section for $t\bar t$
production at International Linear Collider (ILC)
\cite{Weiglein:2004hn} is about three orders less than the LHC, the
very clean environment of the ILC experiments, as well as
polarization of the initial beams make it an attractive platform for
further and complementary investigation of top quarks.

In this work we exploit implications of unparticle physics in the
pair productions of top quarks, and show how the interactions
between unparticles and SM fields can be probed at the LHC with
$\sqrt s=14$ TeV, and at ILC with $\sqrt s$=0.5 TeV by obtaining the
modifications on SM predictions for the $t\bar t$ ($tt$), and $t\bar
t$ production cross sections, respectively. A feature of our work
worth emphasizing is that we analyze the top pair production cross
sections by deriving all the expressions analytically and compute
the contributions of all types of vector, tensor and scalar
unparticles having three different Lorentz structures.

A work which addresses the effects of unparticle physics on top
quark pair production at hadron colliders has recently appeared
while this work was still in progress \cite{Choudhury:2007cq}. The
primary focus in that work is on Tevatron regime. In addition to
this complementary nature in the thematically overlapping part on
hadronic colliders, we have an additional novel part devoted to the
discussion of the same phenomena at ILC. Thus we think these two
works together form a complete set concerning the top pair
production in unparticle physics.

Using the scale invariance, the following propagators for
unparticles with different Lorentz structures can be obtained
\cite{Georgi:2007ek,Cheung:2007ap}:
\begin{eqnarray}\label{pro}
\mathrm{Scalar:}~~\Delta_S&=&\frac{iA_{d_{\mathcal{U}}}}{2\sin(\pi
d_{\mathcal
U})}(-q^2)^{d_{\mathcal U}-2}\nonumber\\
\mathrm{Vector:}~~\Delta_V&=&\frac{iA_{d_{\mathcal{U}}}}{2\sin(\pi
d_{\mathcal
U})}(-q^2)^{d_{\mathcal U}-2}\pi_{\mu\nu}\\
\mathrm{Tensor:}~~\Delta_T&=&\frac{iA_{d_{\mathcal{U}}}}{2\sin(\pi
d_{\mathcal U})}(-q^2)^{d_{\mathcal
U}-2}T_{\mu\nu,\rho\sigma}\nonumber
\end{eqnarray}
where
\begin{eqnarray}
\pi^{\mu\nu}(q) &=& - g^{\mu \nu} + \frac{q^\mu q^\nu }{ q^2}
\;\nonumber\\ T^{\mu\nu,\rho\sigma}(q) &=& \frac{1}{2} \, \left\{
   \pi^{\mu\rho}(q)\  \pi^{\nu\sigma}(q)
        + \pi^{\mu\sigma}(q)\  \pi^{\nu\rho}(q) - \frac{2}{3}\
          \pi^{\mu\nu}(q)\  \pi^{\rho\sigma}(q)  \right\} \;
\end{eqnarray}
and
\begin{equation}
   A_{d_\U}={16\pi^2\sqrt{\pi}\over (2\pi)^{2{d_\U}}}
       { \Gamma({d_\U}+{1\over
       2})\over\Gamma({d_\U}-1)\Gamma(2\,{d_\U})}  \; .
\end{equation}

In the propagators given in Eq.(\ref{pro}) the square of four
momentum transfer through unparticles, $q^2$, has the following
structures:

$\bullet$ $(-q^2)^{d_{\mathcal U}-2}=|q^2|^{d_{\mathcal
U}-2}e^{-id_{\U}\pi}$ in the $s$-channel for which $q^2$ is positive

$\bullet$ $(-q^2)^{d_{\mathcal U}-2}=|q^2|^{d_{\mathcal U}-2}$ in
the $t$ or $u$-channels for which $q^2$ is negative.

 The SM gauge invariant effective interactions of scalar, vector and tensor
unparticles with the SM fields are given by
\cite{Cheung:2007ap,Chen:2007qr};
\begin{eqnarray}
&& \lambda_0 \frac{ 1}{\Lambda_\U^{d_\U-1}} \bar f f O_\U\;, \;\;
\lambda_0 \frac{1}{\Lambda_\U^{d_\U-1} } \bar f i
      \gamma^5 f O_\U\;\;
, \lambda_0 \frac{1}{\Lambda_\U^{d_\U} } G_{\alpha\beta}
G^{\alpha\beta} O_\U \;,
\label{lambda0}\nonumber \\
&&\lambda_1 \frac{1}{\Lambda_\U^{d_\U - 1} }c_v\, \bar f \gamma_\mu
f \, O_\U^\mu \;, \;\; \lambda_1 \frac{1}{\Lambda_\U^{d_\U - 1}
}c_a\, \bar f \gamma_\mu \gamma_5 f \, O_\U^\mu \;,
\label{lambda1} \\
&&- \frac{1}{4}\lambda_2 \frac{1}{\Lambda_\U^{d_\U} } \bar f \,i
   \left(  \gamma_\mu \stackrel{\leftrightarrow}{D}_\nu +
           \gamma_\nu \stackrel{\leftrightarrow}{D}_\mu \right )
  f  \,  O_\U^{\mu\nu} \;,
   \lambda_2 \frac{1}{\Lambda_\U^{d_\U} } G_{\mu\alpha}
G_{\nu}^{\;\alpha} O_\U^{\mu\nu}
 \;,\nonumber
\end{eqnarray}
where $\lambda_i$ ($i=0, 1, 2$) are dimensionless effective
couplings labeling scalar, vector and tensor unparticle operators,
respectively. $c_v, c_a$ represent vector and axial vector couplings
of vector unparticle, respectively. $D_{\mu}$ is the covariant
derivative, $f$ are SM fermions. Finally, $G_{\alpha\beta}$ are the
gluon field strength.

In section II, we will provide analytical expressions of
differential cross sections for $t\bar t$ (and $t t$) productions in
$pp$ collisions. In section III, we provide the corresponding
expressions for $e^+e^-$ collisions. In section IV, we will analyze
the results numerically, for LHC and ILC regimes, and finally
discuss our conclusions.

\section{$t\bar t$ (and $tt$) production in $pp$ collisions with unparticles}
In QCD $t\bar t$ production originates either from quark-antiquark
annihilation or gluon fusion. Below we give the explicit analytical
expressions of the leading order (LO) differential cross sections
for pair productions with the contributions of vector, tensor and
scalar unparticles. We have assumed that unparticles are colorless.

The color and spin averaged partonic differential cross sections for
$t\bar t$ productions via quark-antiquark annihilations are given
by;\\
 \textbf{i)} Vector unparticle
\begin{eqnarray}
\frac{d\hat\sigma}{d\hat t}(q\bar q\rightarrow t\bar
t)&=&\frac{A_V^2}{8\pi\hat s^2(\hat
s)^{4-2d_{\mathcal{U}}}}\Big[c_a^4(2m^4-4(\hat s+\hat t)m^2+(\hat
s+\hat t)^2+\hat t^2)\nonumber\\&+&c_v^4(2m^4-4\hat tm^2+(\hat
s+\hat t)^2+\hat t^2)\nonumber\\&+&2c_v^2c_a^2(2m^4-2(3\hat s+2\hat
t)m^2+3\hat
s^2+2\hat t^2+6\hat s\hat t)\Big]\nonumber\\
&+&\frac{d\sigma_{q\bar q}^0}{d\hat t}
\end{eqnarray}
where
\begin{eqnarray}
A_V=\frac{\lambda_1^2A_{d_{\mathcal{U}}}}{2\sin(d_{\mathcal{U}}\pi)\Lambda^{2(d_{\mathcal{U}}-1)}},
\end{eqnarray}
$\alpha_s$ is the strong coupling constant, $m$ denotes top quark
mass, $\Lambda$ refers to scale $\Lambda_{\U}$ up to which effective
theories valid and $\frac{d\sigma_{q\bar q}^0}{d\hat t}$ is the SM
part:
\begin{eqnarray}
\frac{d\sigma_{q\bar q}^0}{d\hat t}=\frac{4\pi\alpha_s^2}{9\hat
s^4}\Big[2m^4-4\hat tm^2+(\hat s+\hat t)^2+\hat t^2\Big]
\end{eqnarray}
\\
\textbf{ii)} Tensor unparticle
\begin{eqnarray}
\frac{d\hat\sigma}{d\hat t}(q\bar q\rightarrow t\bar
t)&=&\frac{A_T^2}{2\pi\hat s^2(\hat
s)^{4-2d_{\mathcal{U}}}}\Big[32m^8-32(\hat s+4\hat t)m^6+2(5\hat
s^2+64\hat t\hat s+96\hat t^2)m^4\nonumber\\&-&4(\hat s^3+13\hat
t\hat s^2
40\hat t^2\hat s+32\hat t^3)m^2+\hat s^4+32\hat t^4+64\hat s\hat t^3+42\hat s^2\hat t^2+10\hat s^3\hat t\Big]\nonumber\\
&+&\frac{d\sigma_{q\bar q}^0}{d\hat t}
\end{eqnarray}
where
\begin{eqnarray}
A_T=\frac{\lambda_2^2A_{d_{\mathcal{U}}}}{32\sin(
d_{\mathcal{U}}\pi)\Lambda^{2d_{\mathcal{U}}}}
\end{eqnarray}
\\
 \textbf{iii)} Scalar unparticle
\begin{eqnarray}
\frac{d\hat\sigma}{d\hat t}(q\bar q\rightarrow t\bar
t)=\frac{A_S^2}{4\pi\hat s^2(\hat s)^{4-2d_{\mathcal{U}}}}\hat
s(\hat s-2m^2) &+&\frac{d\sigma_{q\bar q}^0}{d\hat t}
\end{eqnarray}
where
\begin{eqnarray}
A_S=\frac{\lambda_0^2}{\lambda_1^2}A_V
\end{eqnarray}
Note that there is no interference contribution in this case.

In unparticle physics there are also Flavor Violating processes
$q\bar q'\rightarrow t\bar t$ ($q,q'=u, c$) via t-channel exchange
of unparticles. Here we have two cases. In the case of $q=q'$ the
s-channel SM reaction is to be included. But the case of $q\neq q'$,
obviously does not have an SM counterpart. Differential cross
sections in the FV case are given below. \\
\textbf{a)} The case $q=q'$ ($q,q'=u, c$):
\begin{eqnarray}\label{13}
\frac{d\hat\sigma^{FV}_V}{d\hat t}(q\bar q\rightarrow t\bar
t)&=&\frac{A_V^2}{16\pi\hat s^2(-\hat
t)^{4-2d_{\mathcal{U}}}}\frac{1}{\hat t^2}\Big[(\tilde c_v^4+\tilde
c_a^4)(m^8-2\hat tm^6+\hat t(4\hat s+3\hat t)m^4-4\hat t^2(2\hat
s+\hat t)m^2\nonumber\\&+&2\hat t^2(2\hat s^2+2\hat t\hat s+\hat
t^2))+2\tilde c_v^2\tilde c_a^2(m^8-2\hat tm^6+\hat t(4\hat s+7\hat
t)m^4-4\hat t^2(2\hat s+3\hat t)m^2\nonumber\\&+&2\hat t^2(2\hat
s^2+6\hat t\hat s+3\hat t^2))
\Big]\nonumber\\
&-&\frac{2\alpha_s A_V}{9\hat s^3(-\hat
t)^{2-d_{\mathcal{U}}}}\frac{1}{\hat t}\Big[(\tilde c_v^2+\tilde c_a^2)(m^6+\hat sm^4-\hat t(2\hat s+3\hat t)m^2+2\hat t(\hat s+\hat t)^2)\Big]\nonumber\\
&+&\frac{d\sigma_{q\bar q}^0}{d\hat t}
\end{eqnarray}
where and $\tilde c_v, \tilde c_a$ are FV vector and axial vector
couplings of vector unparticle, respectively.

\begin{eqnarray}\label{14}
\frac{d\hat\sigma^{FV}_T}{d\hat t}(q\bar q\rightarrow t\bar
t)&=&\frac{ A_T^2}{18\pi\hat s^2(-\hat
t)^{4-2d_{\mathcal{U}}}}\frac{1}{\hat t^4}\Big[8m^{16}-48\hat
tm^{14}+3\hat t(32\hat s+43\hat t)m^{12}\nonumber\\&-&2\hat
t^2(264\hat s+107\hat t)m^{10}+3\hat t^2(128\hat s^2+414\hat t\hat
s+85\hat t^2)m^8\nonumber\\&-&12\hat t^3(124\hat s^2+134\hat t\hat
s+19\hat t^2)m^6+\hat t^3(576\hat s^3+2202\hat t\hat s^2+1212\hat
t^2\hat s+143\hat t^3)m^4\nonumber\\&-&18\hat t^4(64\hat s^3+82\hat
t\hat s^2+28\hat t^2\hat s+3\hat t^3)m^2\nonumber\\&+&9\hat
t^4(32\hat s^4+64\hat t\hat s^3+42\hat t^2\hat s^2+10\hat t^3\hat
s+\hat t^4)\Big]\nonumber\\
&-&\frac{8\alpha_s A_T}{27\hat s^3(-\hat
t)^{2-d_{\mathcal{U}}}}\frac{1}{\hat t^2}\Big[2m^{10}+(2\hat s-5\hat
t)m^8-4\hat s\hat tm^6+2\hat(6\hat s^2+10\hat t\hat s+5\hat
t^2)m^4\nonumber\\&-&2\hat t^2(12\hat s^2+18\hat t\hat s+5\hat
t^2)m^2+3\hat t^2(\hat s+\hat t)^2(4\hat s+\hat
t)\Big]\nonumber\\&+&\frac{d\sigma_{q\bar q}^0}{d\hat t}
\end{eqnarray}

\begin{eqnarray}\label{15}
\frac{d\hat\sigma^{FV}_S}{d\hat t}(q\bar q\rightarrow t\bar
t)&=&\frac{ A_S^2}{4\pi\hat s^2(-\hat
t)^{4-2d_{\mathcal{U}}}}(m^2-\hat t)^2-\frac{4\alpha_s A_S}{9\hat
s^3(-\hat t)^{2-d_{\mathcal{U}}}}(\hat sm^2+(m^2-\hat
t)^2)\nonumber\\&+&\frac{d\sigma_{q\bar q}^0}{d\hat t}
\end{eqnarray}
\\
\textbf{b)} The case $q\neq q'$ ($q,q'=u, c$):

 The corresponding differential cross sections for vector, tensor
 and scalar unparticles would be the terms involving
 $ A_V^2$ only in eq.~(\ref{13}), $A_T^2$ in eq.~(\ref{14}) and $ A_S^2$ in eq.~(\ref{15}),
 respectively, which are obtained by disregarding the SM parts and
 interferences between the unparticle and SM sectors.

After completing quark-antiquark annihilations, we now analyze
$t\bar t$ productions originating from \textbf{gluon fusion}. In
this case, to produce $t\bar t$ we have four Feynman diagrams in two
cases of tensor and scalar unparticle exchanges, as the vector
unparticle do not couple to gluons. One out of these four is
$s$-channel exchange of unparticle and the other three are SM
($s$,$t$ and $u$ channel) diagrams. The diagram mediated by
unparticles do not interfere with that of SM model $s$-channel
mediated by gluon, and interfere only with $t$ and $u$ channel
processes mediated by quarks. Hence, the differential cross sections
for the process $gg\rightarrow t\bar t$ have the following forms:

\textbf{i)} Tensor unparticle\\
\begin{eqnarray}
\frac{d\hat\sigma}{d\hat t}(gg\rightarrow t\bar
t)&=&\frac{3A_T^2}{\pi\hat s^2(\hat
s)^{4-2d_{\mathcal{U}}}}\Big[-2m^8+8\hat t m^6-(\hat s^2+4\hat t\hat
s+12 \hat t^2)m^4+2\hat t(\hat s+\hat t)^2m^2\nonumber\\&-&\hat
t(\hat s^3+3\hat t \hat s^2+4\hat t^2\hat s+2\hat t^3)\Big]\nonumber\\
&+&\frac{\alpha_sA_T\cos(d_\U\pi)}{12\hat s^2(\hat
s)^{2-d_{\mathcal{U}}}}\Big[\frac{1}{\hat t-m^2}(2m^6-2(2\hat
s+5\hat t)m^4-(3\hat s^2+2\hat t \hat s-14\hat
t^2)m^2\nonumber\\&-&3\hat t(\hat s^2+2\hat t\hat s+2\hat
t^2))\nonumber\\&+&\frac{1}{\hat u-m^2}(2m^6-2(2\hat s+5\hat
u)m^4-(3\hat s^2+2\hat u \hat s-14\hat u^2)m^2\nonumber\\&-&3\hat
u(\hat s^2+2\hat u\hat s+2\hat
u^2))\Big]+\frac{d\sigma_{gg}^0}{d\hat t}
\end{eqnarray}
where $\frac{d\sigma_{gg}^0}{d\hat t}$ is SM part of the
differential cross section of the gluon fusion in $t\bar t$
production, and is given by
\begin{eqnarray}
\frac{d\sigma_{gg}^0}{d\hat t}&=&\frac{\pi\alpha_s^2}{\hat
s^2}\Big[\frac{3}{4\hat s^2}(m^2-\hat t)(m^2-\hat u)-\frac{1}{2(\hat
t-m^2)^2}(3m^4-\hat sm^2+\hat t(\hat s+\hat
t))\nonumber\\&-&\frac{1}{2(\hat u-m^2)^2}(3m^4-\hat sm^2+\hat
u(\hat s+\hat u))-\frac{3}{8\hat s(\hat t-m^2)}(3m^4-2(\hat s+2\hat
t)m^2+\hat t(\hat s+\hat t))\nonumber\\&-&\frac{3}{8\hat s(\hat
u-m^2)}(3m^4-2(\hat s+2\hat u)m^2+\hat u(\hat s+\hat
u))\nonumber\\&-&\frac{1}{24(\hat t-m^2)(\hat u-m^2)}(m^2(2m^2-\hat
t+\hat u))\Big]
\end{eqnarray}

ii) Scalar unparticle\\
\begin{eqnarray}
\frac{d\hat\sigma}{d\hat t}(gg\rightarrow t\bar
t)&=&\frac{3A_S'^2}{256\pi\hat s^2(\hat
s)^{4-2d_{\mathcal{U}}}}\Big[\hat s^2(2\hat s-2m^2)\Big]\nonumber\\
&+&\frac{\alpha_sA_S'\cos(d_\U\pi)}{64\hat s^2(\hat
s)^{2-d_{\mathcal{U}}}}\Big[\frac{1}{\hat t-m^2}m(2m^4+(5\hat
s-4\hat t)m^2-2\hat s^2+2\hat t^2-\hat s\hat
t)\nonumber\\&+&\frac{1}{\hat u-m^2}m(2m^4+(5\hat s-4\hat
u)m^2-2\hat s^2+2\hat u^2-\hat s\hat u)\Big]\nonumber\\
&+&\frac{d\sigma_{gg}^0}{d\hat t}
\end{eqnarray}
where
\begin{eqnarray}
A_S'=\frac{2\lambda_0^2A_{d_{\mathcal{U}}}}{\sin(d_{\mathcal{U}}\pi)\Lambda^{2d_{\mathcal{U}}-1}}
\end{eqnarray}

Finally, the cross sections for the processes $q q'\rightarrow t t$
($q, q'=u, c$) for the FV case in unparticle physics are:

\textbf{i)} Vector unparticle\\
\begin{eqnarray}
\frac{d\hat\sigma}{d\hat t}(q q'\rightarrow t
t)&=&\frac{A_V^2}{16\pi\hat s^2}\Bigg\{\Big\{\Big[\frac{1}{(-\hat
t)^{4-2d_{\mathcal{U}}}}\frac{1}{\hat t^2}\Big((\tilde c_v^4+\tilde
c_a^4)(m^8-2\hat tm^6+\hat t(4\hat s+3\hat t)m^4-4\hat t^2(2\hat
s+\hat t)m^2\nonumber\\&+&2\hat t^2(2\hat s^2+2\hat t\hat s+\hat
t^2))+2\tilde c_v^2\tilde c_a^2(m^8-2\hat tm^6+\hat t(4\hat s-\hat
t)m^4+4\hat t^2(\hat t-2\hat s)m^2\nonumber\\&-&2\hat t^2(\hat
t^2+2\hat t\hat s-2\hat s^2))\Big)\Big] +[\hat t\leftrightarrow\hat
u]\Big\}+\frac{1}{6(-\hat t)^{2-d_\U}(-\hat
u)^{2-d_\U}}\frac{1}{\hat t\hat u}\Big[(\tilde c_v^4+\tilde
c_a^4)(3m^8 \nonumber\\&-&(\hat t+\hat u)(3m^6-6m^2\hat t\hat
u+4\hat t\hat u(\hat t+\hat u))+m^4(2\hat t^2+3\hat t\hat u+2\hat
u^2)) +2\tilde c_v^2\tilde c_a^2(m^8\nonumber\\&-&(\hat t+\hat
u)(13m^6-26m^2\hat t\hat u+12\hat t\hat u(\hat t+\hat u))+m^4(6\hat
t^2+9\hat t\hat u+6\hat u^2)\Big]\Bigg\}
\end{eqnarray}

\textbf{ii)} Tensor unparticle\\
\begin{eqnarray}
\frac{d\hat\sigma}{d\hat t}(q q'\rightarrow t t)&=&\frac{
A_T^2}{18\pi\hat s^2}\Bigg\{\Big\{\Big[\frac{1}{(-\hat
t)^{4-2d_{\mathcal{U}}}}\frac{1}{\hat t^4}\Big(8m^{16}-48\hat
tm^{14}+3\hat t(32\hat s+43\hat t)m^{12}-2\hat t^2(264\hat
s\nonumber\\&+&107\hat t)m^{10}+3\hat t^2(128\hat s^2+414\hat t\hat
s+85\hat t^2)m^8-12\hat t^3(124\hat s^2+134\hat t\hat s+19\hat
t^2)m^6\nonumber\\&+&\hat t^3(576\hat s^3+2202\hat t\hat
s^2+1212\hat t^2\hat s+143\hat t^3)m^4-18\hat t^4(64\hat s^3+82\hat
t\hat s^2+28\hat t^2\hat s\nonumber\\&+&3\hat t^3)m^2+9\hat
t^4(32\hat s^4+64\hat t\hat s^3+42\hat t^2\hat s^2+10\hat t^3\hat
s+\hat t^4)\Big)\Big]+[\hat t\leftrightarrow\hat
u]\Big\}\nonumber\\&+&\frac{1}{(-\hat t)^{2-d_{\mathcal{U}}}(-\hat
u)^{2-d_{\mathcal{U}}}}\frac{1}{3\hat u^2\hat
t^2}\Big[12m^{16}+26m^{14}(\hat t+\hat u) +m^{12}(-28\hat t^2+6\hat
t\hat u\nonumber\\&-&28\hat u^2)-9\hat t^2\hat u^2(\hat t+\hat
u)^2(4\hat t^2+17\hat t\hat u+4\hat t^2)+m^{10}(20\hat t^3 -27\hat
t^2\hat u-27\hat t\hat u^2+20\hat u^3)\nonumber\\&-&4m^6\hat t\hat
u(36\hat t^3+149\hat t^2\hat u+149\hat t\hat u^2+36\hat
u^3)+3m^2\hat t^2\hat u^2(52\hat t^3 +231\hat t^2\hat u+231\hat
t\hat u^2\nonumber\\&+&52\hat u^3)+m^8(-6\hat t^4+136\hat t^3\hat
u+384\hat t^2\hat u^2+136\hat t\hat u^3-6\hat u^4) +m^4\hat t\hat
u(36\hat t^4+45\hat t^3\hat u\nonumber\\&-&124\hat t^2\hat
u^2+45\hat t\hat u^3+36\hat u^4)\Big]\Bigg\}
\end{eqnarray}

\textbf{iii)} Scalar unparticle\\
\begin{eqnarray}
\frac{d\hat\sigma}{d\hat t}(q q'\rightarrow t
t)&=&\frac{A_S^2}{4\pi\hat s^2}\Bigg\{\Big\{\Big[\frac{1}{(-\hat
t)^{4-2d_{\mathcal{U}}}}(m^2-\hat t)^2\Big]+[\hat
t\leftrightarrow\hat u]\Big\}\nonumber\\&+&\frac{1}{6(-\hat
t)^{2-d_{\mathcal{U}}}(-\hat u)^{2-d_{\mathcal{U}}}}(m^4+m^2(\hat
s-2\hat t)+\hat t(\hat s+\hat t))\Bigg\}
\end{eqnarray}

As we are finished with the construction of the analytical
expressions for the partonic differential cross sections for the
$t\bar t$ and $tt$ productions in $pp$ collisions now we will derive
the corresponding expressions for the process $e^+e^-\rightarrow
t\bar t$ in the next section.
\section{$t\bar t$ production in $e^+e^-$ collisions with unparticles}
In this section we present the differential cross sections for the
processes $e^+e^-\rightarrow t\bar t$ which occur via the s-channel
exchanges of unparticles and of usual electroweak bosons, $\gamma$
and Z\\
\textbf{i)} Vector unparticle\\
\begin{eqnarray}\label{eq22}
\frac{d\hat\sigma}{d\hat t}(e^+e^-\rightarrow t\bar
t)&=&\frac{3A_V^2}{8\pi\hat s^2(\hat
s)^{4-2d_{\mathcal{U}}}}\Big[c_a^4(2m^4-4(\hat s+\hat t)m^2+(\hat
s+\hat t)^2+\hat t^2)+c_v^4(2m^4-4\hat tm^2\nonumber\\&+&(\hat
s+\hat t)^2+\hat t^2)+2c_v^2c_a^2(2m^4-2(3\hat s+2\hat t)m^2+3\hat
s^2+2\hat t^2+6\hat s\hat t)\Big]\nonumber\\&+&\frac{3A_V\alpha
e_q\cos(d_\U\pi)}{\hat s^3(\hat s)^{2-d_\U}}\Big[c_a^2\hat s(\hat t
-\hat u)+c_v^2(2m^4+2\hat sm^2-2\hat t\hat u-\hat s(\hat t+\hat
u)\Big]\nonumber\\&+&\frac{3\alpha A_V[(\hat s-M_Z^2)\cos(\pi d_\U)
+\Gamma_ZM_Z\sin(\pi d_\U)]}{2c_W^2s_W^2\hat s^3(\hat s)^{2-d_\U}[(\hat s-M_Z^2)^2+\Gamma_Z^2M_Z^2]}\Big[v_ev_t(\hat t-\hat u)\hat s^2+a_ea_tc_a^2(8m^6\nonumber\\
&-&2(5\hat s +4(\hat t+\hat u))m^4+2(\hat s^2+3\hat s(\hat t+\hat u)+(\hat t+\hat u)^2)m^2-\hat s(2\hat t\hat u+\hat s(\hat t+\hat u)))\nonumber\\
&-&2c_vc_a(a_ev_t\hat s
(m^4+\hat sm^2-\hat u(\hat s+\hat t))+v_ea_t(4m^6-(5\hat s+(\hat t+\hat u))m^4+(\hat s^2\nonumber\\
&+&3\hat s(\hat t+\hat u)+(\hat t+\hat u)^2)m^2-\hat s\hat u(\hat s+\hat t)))+c_v^2\hat s(a_ea_t\hat s(\hat t-\hat u)+v_ev_t(2m^4\nonumber\\
&+&2\hat sm^2-2\hat t\hat u
-\hat s(\hat t+\hat u)))\Big]+\frac{d\sigma^0}{d\hat t}
\end{eqnarray}
where $s_W=\sin\theta_W$ and $c_W=\cos\theta_W$ with $\theta_W$
being Weinberg angle and $d\hat\sigma^0/d\hat t$ is the SM part
given by
\begin{eqnarray}\label{eq23}
\frac{d\sigma^0}{d\hat t}&=&\frac{6\pi\alpha^2}{\hat
s^2}\Bigg\{\frac{e_q^2}{\hat s^2}[2m^4+2\hat sm^2-2\hat t\hat u-\hat
s(\hat t+\hat u)]\nonumber\\&+&\frac{1}{4c_W^4s_W^4[(\hat
s-M_Z^2)^2+\Gamma_Z^2M_Z^2]} \Big[4a_ev_ea_tv_t\hat s(\hat s+2\hat
t-3m^2)+(v_e^2+a_e^2)(v_t^2+a_t^2)[2m^4\nonumber\\&+&\hat s^2+2\hat
s\hat t +2\hat t^2-4m^2\hat t]-4m^2\hat s
a_t^2(a_e^2+v_e^2)\Big]+\frac{e_q(\hat s-M_Z^2)}{\hat s
c_W^2s_W^2[(\hat s-M_Z^2)^2 +\Gamma_Z^2M_Z^2]}[a_ea_t\hat s(\hat
t-\hat u)\nonumber\\&+&v_ev_t(2m^4+2\hat sm^2-2\hat t\hat u-\hat
s(\hat t+\hat u))]\Bigg\}
\end{eqnarray}
In Eqs.~(\ref{eq22}) and (\ref{eq23}) $v_e$, $a_e$ and $v_t$, $a_t$
are vector and axial vector couplings of Z to electron and top
quark, respectively. We should note that vector unparticle interfere
with photon and Z boson\\
\textbf{ii)} Tensor unparticle\\
\begin{eqnarray}
\frac{d\hat\sigma}{d\hat t}(e^+e^-\rightarrow t\bar
t)&=&\frac{3A_T^2}{2\pi \hat s^2(\hat
s)^{4-2d_\U}}\Big[32m^8-32(\hat s+4\hat t)m^6+2(5\hat s^2+64\hat t
\hat s+96\hat t^2)m^4\nonumber\\&-&4(\hat s^3+13\hat t\hat
s^2+40\hat t^2\hat s+32\hat t^3)m^2+\hat s^4+32\hat t^4+64\hat s\hat
t^3+42\hat s^2\hat t^2 +10\hat s^3\hat
t\Big]\nonumber\\&-&\frac{6A_T\alpha e_q\cos(d_\U\pi)}{\hat s^3(\hat
s)^{2-d_\U}}\Big[8m^6-4(\hat s+6\hat t)m^4+2(\hat s^2+8\hat t\hat
s+12\hat t^2) m^2-(\hat s+2\hat t)^3\Big]\nonumber\\&+&\frac{3\alpha
A_T[(\hat s-M_Z^2)\cos(\pi d_\U) +\Gamma_ZM_Z\sin(\pi
d_\U)]}{2c_W^2s_W^2\hat s^2(\hat s)^{2-d_\U}[(\hat
s-M_Z^2)^2+\Gamma_Z^2M_Z^2]}\Big[a_ea_t\hat s(4\hat sm^2-\hat
s^2+3(\hat t -\hat u)^2)\nonumber\\&+&v_ev_t(\hat t-\hat
u)(4m^4+(8\hat s-4(\hat t+\hat u))m^2-\hat s^2+3(\hat t^2+\hat
u^2)-2\hat t\hat u)\Big]+\frac{d\sigma^0}{d\hat t}
\end{eqnarray}
Here we again have the interference terms as in the vector
unparticle case\\
\textbf{iii)} Scalar unparticle\\
\begin{eqnarray}\label{eq25}
\frac{d\hat\sigma}{d\hat t}(e^+e^-\rightarrow t\bar
t)&=&\frac{3A_S^2}{4\pi\hat s^2(\hat s)^{4-2d_{\mathcal{U}}}}\hat
s(\hat s-2m^2)+\frac{d\sigma^0}{d\hat t}
\end{eqnarray}
It is seen from Eq.~(\ref{eq25}) that the scalar unparticle
contribution does not interfere with those of the neutral
electroweak bosons, $\gamma$ and Z.

Now we are ready to turn our attention to the total cross sections
at the LHC and ILC energies. We will carry out this analysis
numerically in the next section.
\section{Numerical Analysis, Discussions and Conclusions}

In this section, we give the numerical study of top pair production
in unparticle physics at the LHC with $\sqrt s$= 14 TeV and ILC with
$\sqrt s$= 0.5 TeV. The total hadronic cross section for
$pp\rightarrow t\bar t (tt)+X$ is obtained by integrating subprocess
cross sections ,which were calculated in the previous section, over
the parton distribution functions $f_i(x,Q^2)$ that represent the
number density of the parton $i$ carrying the fraction $x$ of the
longitudinal proton momentum:
\begin{eqnarray}
\label{cross1} \sigma\left({ p\, p \rightarrow t\,\bar t (tt)+
X}\right) &=& \int_{\frac{4 m^2}{s}}^{1} d \tau \int_{\tau}^{1}
\frac{d x}{x} \frac{1}{1+\delta_{ij}}\sum_{ij}\Bigg\{ f_i\left(x,
Q^2\right) f_j\left(\frac{\tau}{x}, Q^2\right)\nonumber\\&+&
f_i\left(\frac{\tau}{x}, Q^2\right) f_j\left(x,
Q^2\right)\Bigg\}\hat\sigma\left(\hat s,m \right)
\end{eqnarray}
where $(i,j)=(g,g), (q_{\alpha},\bar q_{\alpha'}),
(q_{\alpha},q_{\alpha'})$; $(\alpha,\alpha')=u, d, s, c, b$. The
factorization scale has been taken as $Q=m$. For the parton
distribution functions, CTEQ5 parametrization \cite{CTQ5} has been
used in numerical calculations.

\begin{figure*}
  \includegraphics[width=10cm]{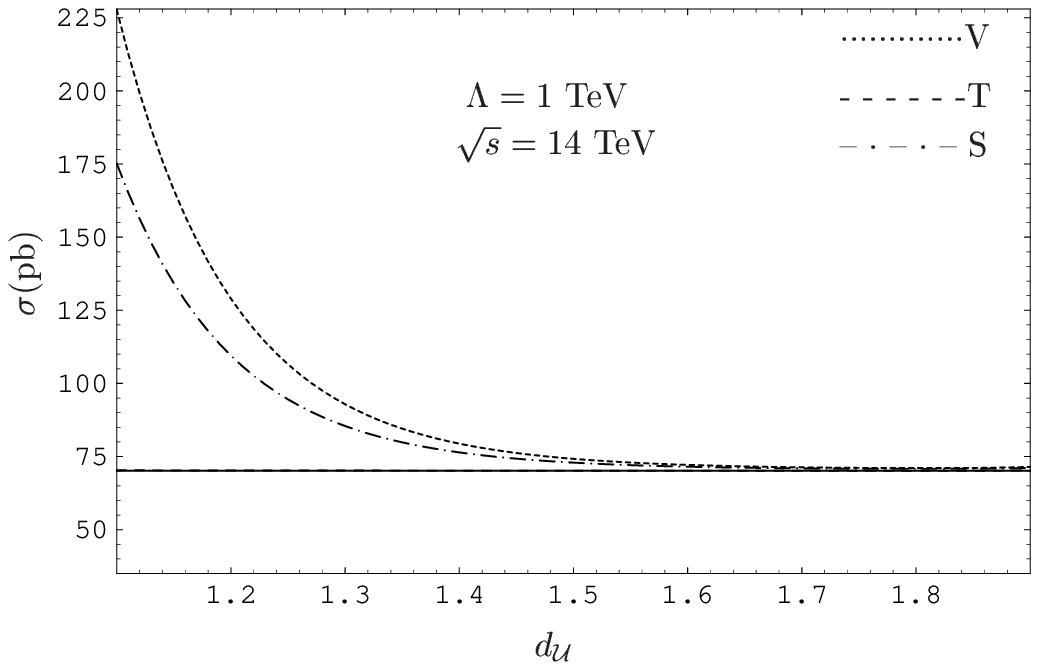}\\
 \caption{Total cross section in pb originating from the reaction $q\bar q\rightarrow t\bar t$ for $\Lambda$= 1 TeV, $\lambda_0=\lambda_1=\lambda_2=1$,
 and $c_v=c_a=1$ at LHC. Solid line corresponds to SM prediction at LO.}\label{fig1}
\end{figure*}

 In Fig.\ref{fig1}, we present the total cross section for
the process $q\bar q\rightarrow t\bar t$ by setting $\Lambda$= 1TeV,
$\lambda_0=\lambda_1=\lambda_2=1$ and $c_v=c_a=1$ at LHC regime with
the CM energies of 14 TeV for the flavor conserving case. From this
figure we see that, main contribution comes from the vector
unparticle, and it enhances the SM result substantially for values
of $d_{\U}$, $1<d_{\U}<1.5$. The tensor contribution, however, is
tiny as expected and insensitive to variations of scale dimension.
It is 70.30 pb for $d_{\U}=1.1$ and 70.20 pb for $d_{\U}=1.9$.
Hence, the contribution of tensor unparticle does not appear in the
Fig.\ref{fig1}. The scalar contribution is quite large and
comparable with that of vector unparticle.

\begin{figure*}
  \includegraphics[width=10cm]{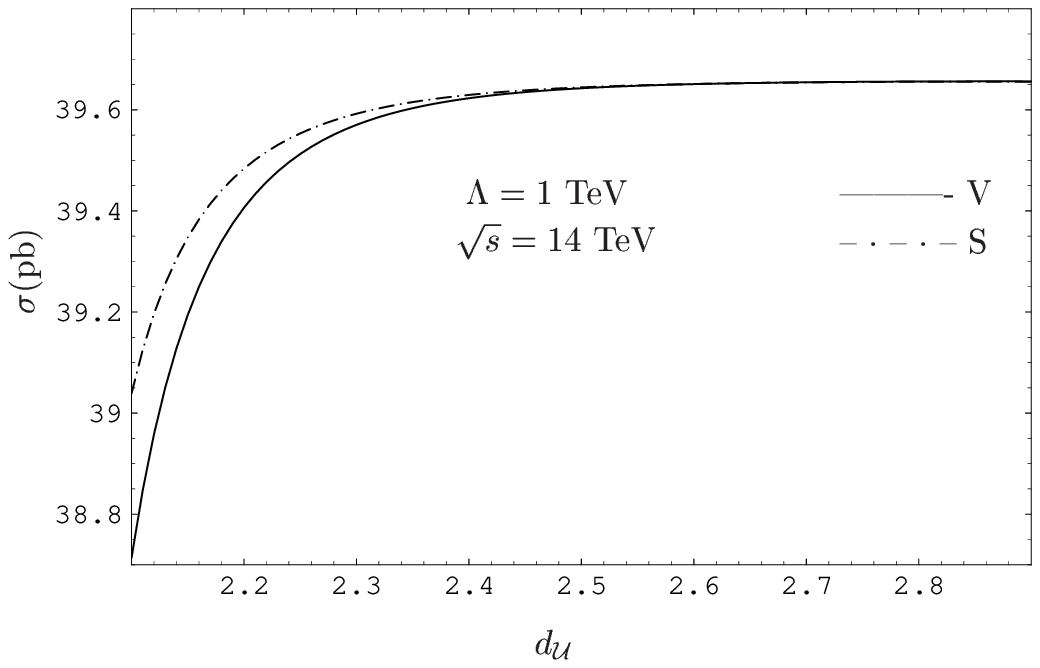}\\
 \caption{Total cross section in pb originating from the flavor violating reaction $q\bar q\rightarrow t\bar t$
($q=u,c$) for $\Lambda$= 1 TeV, $\lambda_0=\lambda_1=1$,
 and $\tilde c_v=\tilde c_a=1$ at LHC.}\label{fig2}
\end{figure*}

\begin{figure*}
  \includegraphics[width=10cm]{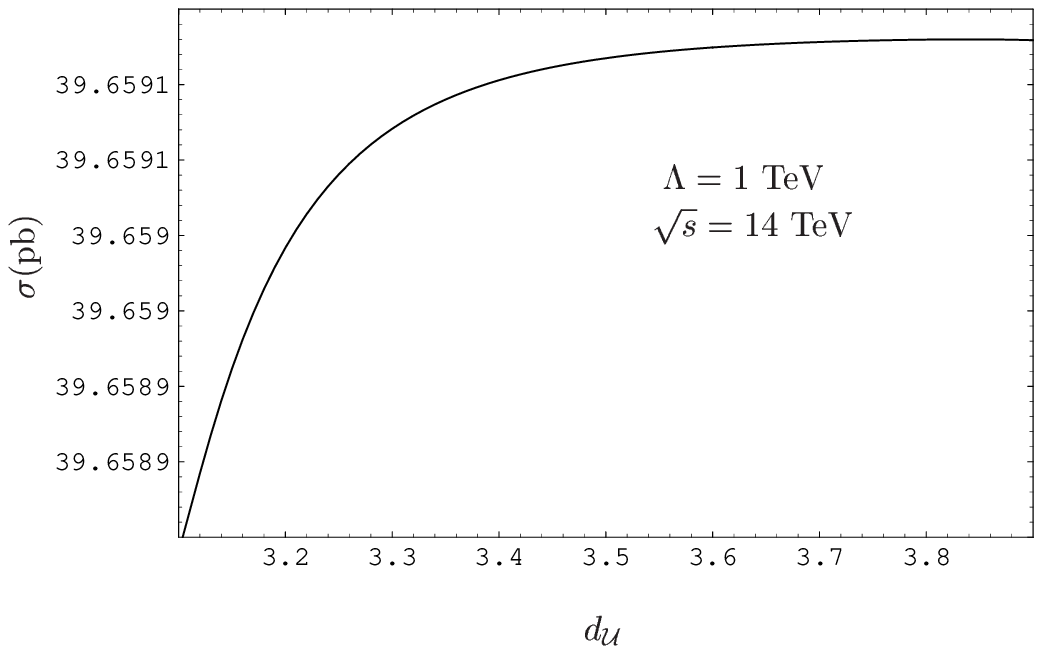}\\
 \caption{Total cross section in pb originating from the flavor violating reaction $q\bar q\rightarrow t\bar t$
($q=u,c$) for $\Lambda$= 1 TeV, $\lambda_2=1$ at LHC through tensor
unparticle}\label{fig3}
\end{figure*}

It has been shown that there are constraints imposed on the
parameters of unparticle physics stemming from some physical
processes like $\mu$ decay \cite{Choudhury:2007js}. That is, if the
coupling strength of the vector unparticle, $\tilde c_v$ and $\tilde
c_a$, for the FV case are chosen to be of the same magnitude as in
the FC case then, $d_{\U}$ must be bounded from below namely,
$d_{\U}>2$ for vector unparticle and $d_{\U}>3$ for the tensor
unparticle, without any constraint on $d_{\U}$ for scalar
unparticle. The results are depicted in Fig.~\ref{fig2} for vector
and scalar contributions, and in Fig.~\ref{fig3} for tensor
contribution. These contributions are almost insensitive to the
variation of $d_{\U}$ beyond certain critical values and is about 40
pb most of which comes from s-channel SM diagram.

\begin{figure*}
  \includegraphics[width=10cm]{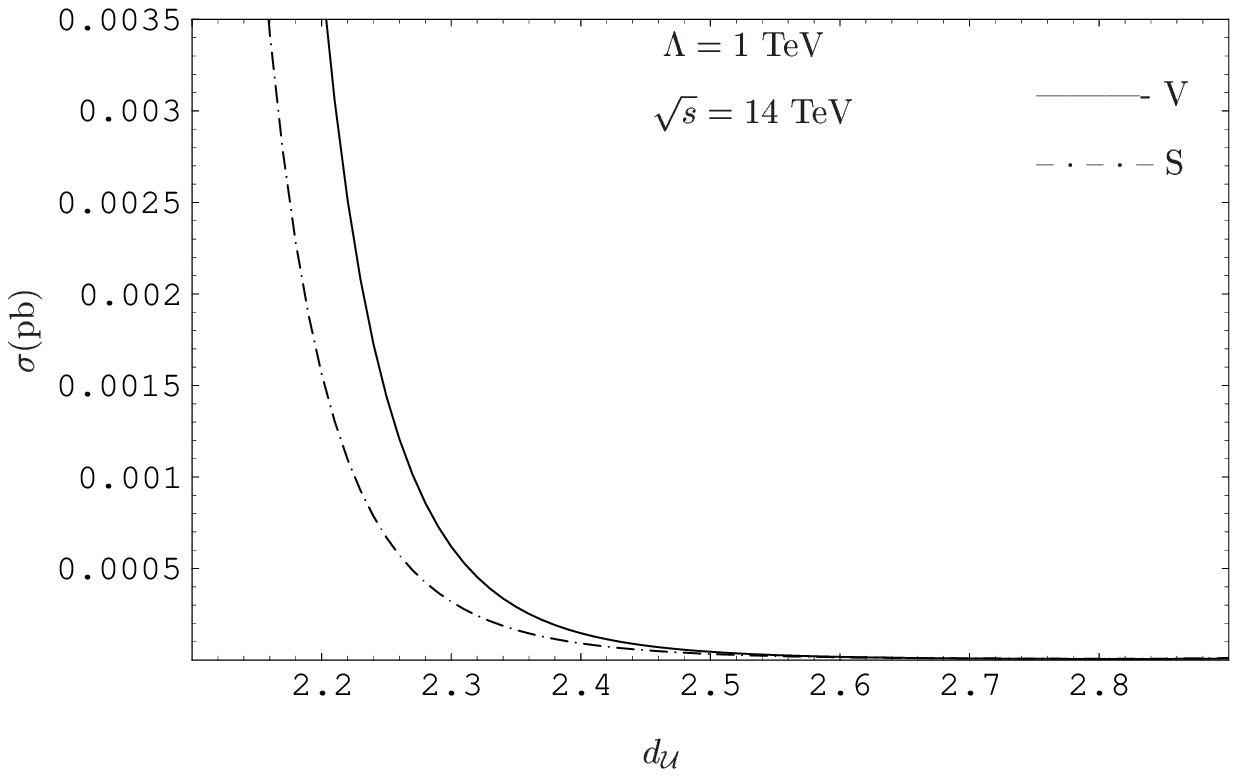}\\
 \caption{Total cross section in pb originating from the flavor violating reaction $q\bar{ q'}\rightarrow t\bar t$
  ($q,q'=u,c$ $q\neq q'$) for $\Lambda$= 1 TeV, $\lambda_0=\lambda_1=1$,
 and $\tilde c_v=\tilde c_a=1$ at LHC.}\label{fig4}
\end{figure*}

\begin{figure*}
  \includegraphics[width=10cm]{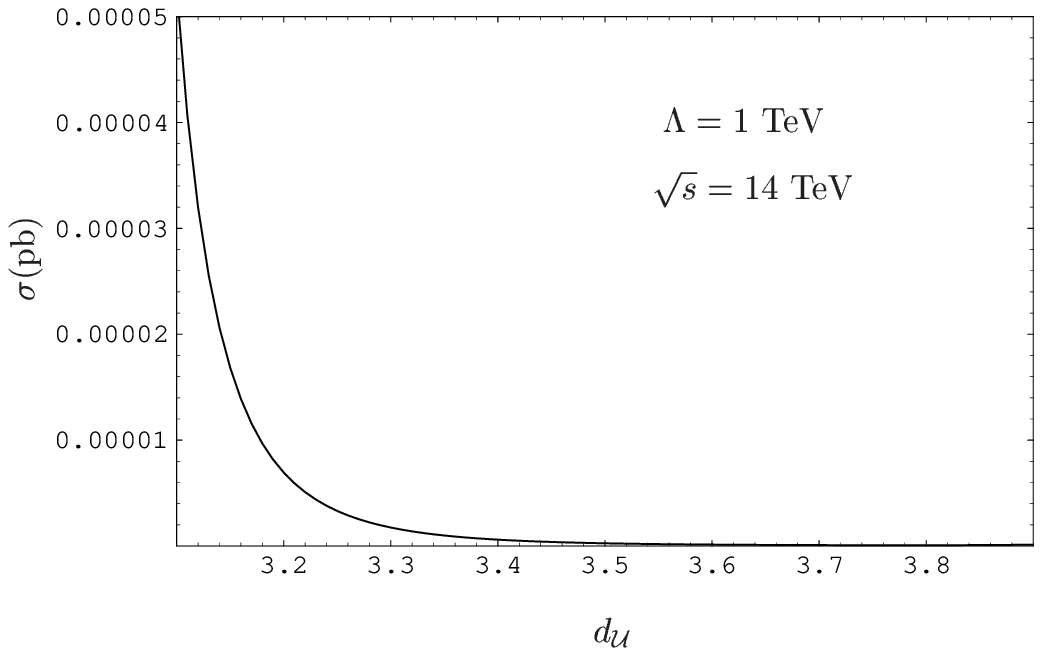}\\
 \caption{Total cross section in pb originating from the flavor violating reaction $q\bar{ q'}\rightarrow t\bar t$
  ($q,q'=u,c$ $q\neq q'$) for $\Lambda$= 1 TeV, $\lambda_2=1$ at LHC through tensor unparticle}\label{fig5}
\end{figure*}

Our results for the $q\bar q'\rightarrow t\bar t$ are depicted in
Figs.~\ref{fig4} and \ref{fig5}, for the vector and scalar, and the
tensor unparticles, respectively, by taking into account the
constraints on $d_{\U}$ as mentioned above. It is clear from these
figures that these results are also very small.

\begin{figure*}
  \includegraphics[width=10cm]{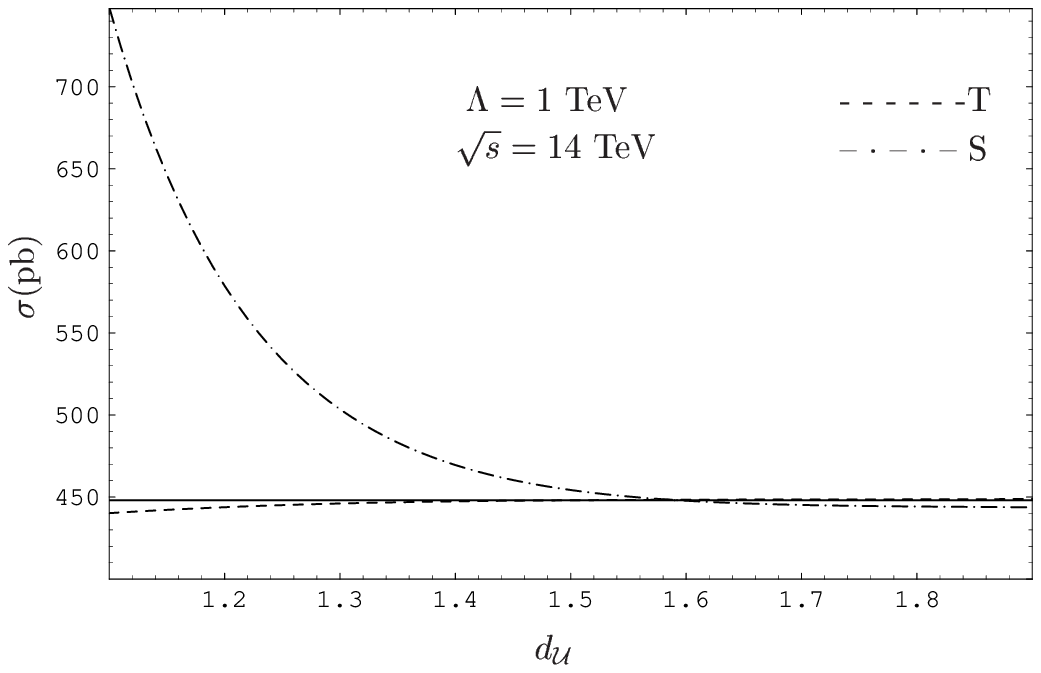}\\
 \caption{Total cross section in pb originating from the reaction
 $gg\rightarrow t\bar t$ for $\Lambda$= 1 TeV, $\lambda_0=\lambda_2=1$ at LHC.  Solid line corresponds to SM prediction at LO.}\label{fig6}
\end{figure*}

In Fig.~\ref{fig6} we plot the dependence of total cross section for
the process $gg\rightarrow t\bar t$ at $\Lambda$=1 TeV,
$\lambda_0=\lambda_1=1$ at LHC energies. Notice that this is a FC
process. As we have already mentioned in the previous section, the
vector unparticle does not give any contribution for this channel.
In this case the main contribution comes from the scalar unparticle;
the enhancement due to this contribution is quite large especially
for $d_{\U}<1.5$ and becomes slightly negative for $d_{\U}>1.6$. The
result for $d_{\U}=1.9$ is 433.8 pb as the SM value at LO is 448 pb.
The contribution of tensor unparticle varies from 440 pb for
$d_{\U}=1.1$ to 448.9 pb for $d_{\U}=1.9$.

\begin{figure*}
  \includegraphics[width=12cm]{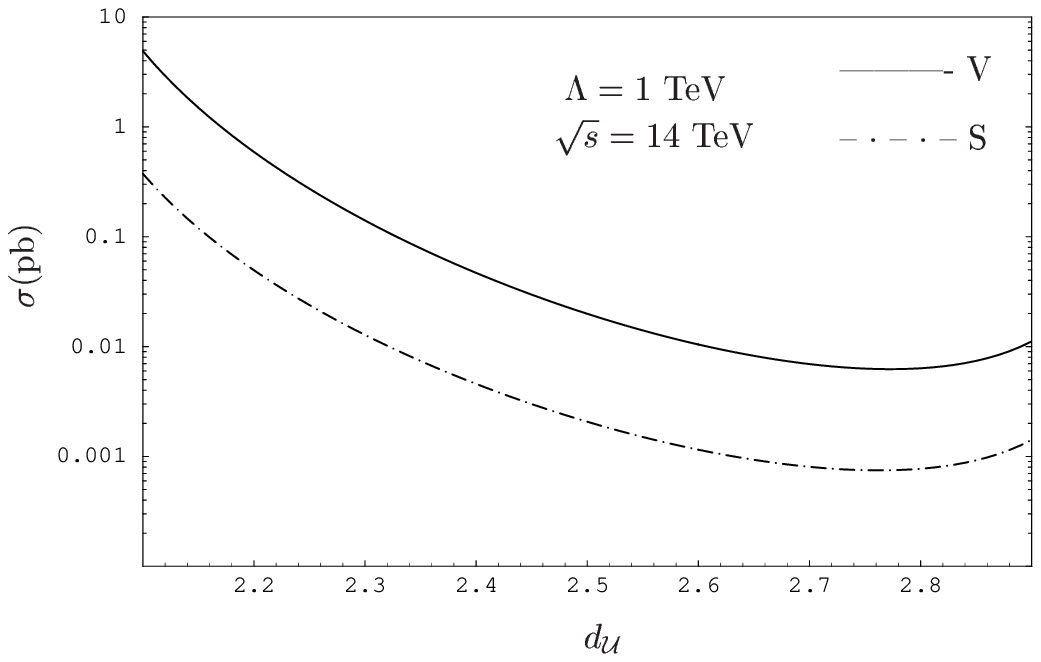}\\
  \caption{Total cross section in pb for ditop production originating from the flavor
   violating reaction $q q\rightarrow t t$ ($q=u,c$) for $\Lambda$= 1 TeV, $\lambda_0=\lambda_1=1$,
 and $\tilde c_v=\tilde c_a=1$ at LHC.}\label{fig7}
\end{figure*}

\begin{figure*}
  \includegraphics[width=12cm]{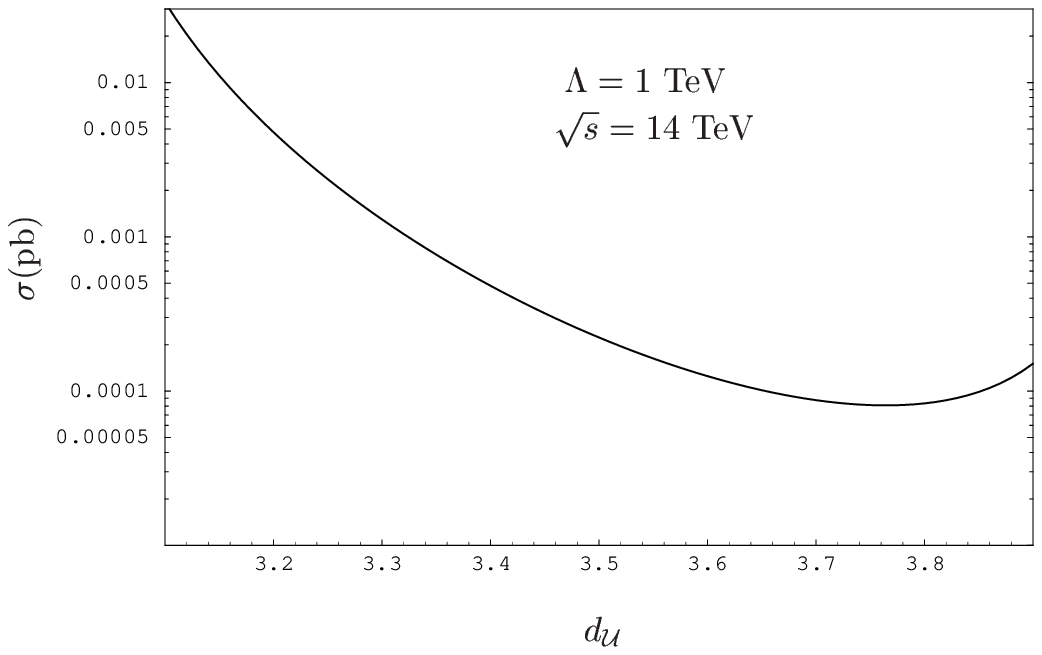}\\
  \caption{Total cross section in pb for ditop production originating from the flavor
   violating reaction $q q\rightarrow t t$ ($q=u,c$) for $\Lambda$= 1 TeV, $\lambda_2=1$ at LHC through tensor unparticle}\label{fig8}
\end{figure*}

In Figs.~\ref{fig7}, \ref{fig8} we plot the cross sections for
$qq\rightarrow tt$ $(q=u,c)$ process for the scalar, vector and the
tensor unparticle contributions, respectively, for the relevant
$d_{\U}$ intervals. The cross section for these FV processes, for
which there is no SM counterpart, turn out to be very small. For the
range of $2<d_{\U}<3$ total cross sections for $tt$ production take
values from 4.9 pb to 0.01 pb for vector unparticle exchange and
from 0.37 pb to 0.0014 pb for scalar unparticle exchange as shown in
Fig.~\ref{fig7}. The result for tensor unparticle exchange varies
between 0.11 and 0.00005 pb for the range of $3<d_{\U}<4$ as shown
in Fig.~\ref{fig7}.

\begin{figure*}
  \includegraphics[width=10cm]{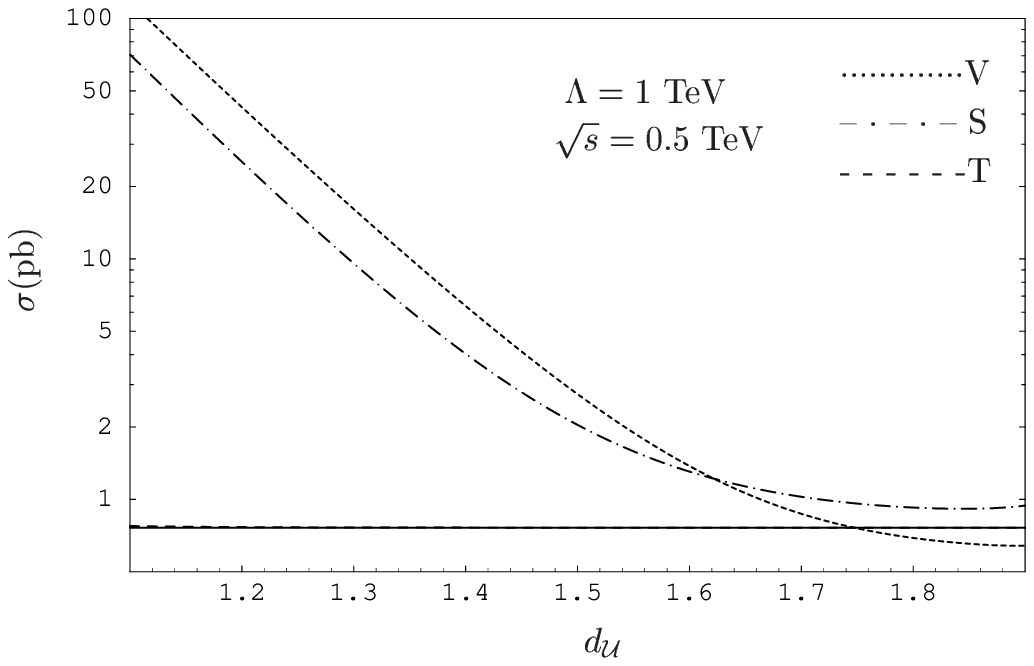}
  \caption{Total cross section in pb for the reaction
 $e^+e^-\rightarrow t\bar t$ at ILC ($\sqrt s$=0.5 TeV). We have set $\Lambda$= 1 TeV, $\lambda_0=\lambda_1=\lambda_2=1$ and $c_v=c_a=1$.
 Solid line corresponds to SM prediction at LO.}\label{fig9}
\end{figure*}

Turning to the case of $t\bar t$ production at the ILC, we see from
Fig.~\ref{fig9} that virtual effects of vector and scalar
unparticles are substantial for specifically the range of
$1<d_{\U}<1.5$. As the SM value is about 0.76 pb, these
contributions are about 116 and 71 pb for $d_{\U}=1.1$,
respectively. Contribution of the tensor unparticle is very small,
however. It changes between 0.774 and 0.762 pb for the range of
$1<d_{\U}<2$.

Effects of unparticle physics can also be investigated on single
productions of top quarks via FV processes which have no
counterparts in the SM. These processes have been investigated for
different types of colliders, namely, lepton-hadron, and linear
$e^+e^-$ colliders, as well as LHC in ref.\cite{Alan:2007ui}. This
study reveals, as we have expected, that there could be observable
effects of unparticles, with rather striking signals for certain
values of $d_\U$, which may provide further insight on unparticle
physics.

In summary we have explored the phenomenology of unparticle physics
with the top quark pair productions in $pp$ and $e^+e^-$ collisions.
Depending on relevant ranges of the scale dimension $d_\U$ we have
found significant enhancements as compared to the SM predictions for
the total cross sections. For the case like-sign top pair production
at LHC our numerical result agrees with that of
\cite{Choudhury:2007cq}, in the case of vector unparticle. Our
results for ILC are rather striking as compared to the LHC results
because, for smaller values of $d_\U$, the enhancements of the cross
sections exceed SM predictions by two orders of magnitude, while the
corresponding enhancements for the LHC case are only about a factor
of three at the most.

\begin{acknowledgements}
This work is partially supported by Abant Izzet Baysal University Research Fund.
\end{acknowledgements}

\newpage


\end{document}